\documentclass[prd,aps,
twocolumn,superscriptaddress,preprintnumbers,nofootinbib,english]{revtex4-1}
\usepackage{pslatex}
\usepackage[pdftex]{graphicx}
\usepackage{psfrag}
\usepackage{epsfig}
\usepackage{color}
\usepackage{cancel}
\usepackage{slashed}
\usepackage{amssymb}
\usepackage{amsmath}
\usepackage{hyperref}

\begin{document}
\title{The EDGES signal: An imprint from the mirror world?}

\author{D. Aristizabal Sierra}%
\affiliation{Universidad T\'ecnica 
  Federico Santa Mar\'{i}a - Departamento de F\'{i}sica\\
  Casilla 110-V, Avda. Espa\~na 1680, Valpara\'{i}so, Chile}
\affiliation{IFPA, Dep. AGO, Universit\'e de Li\`ege, Bat B5, Sart
  Tilman B-4000 Li\`ege 1, Belgium}
\author{Chee Sheng Fong}%
\affiliation{Departamento de F\'isica, Pontif\'icia Universidade Cat\'olica do Rio de Janeiro, 
Rio de Janeiro, Brazil}
\affiliation{Instituto de F\'isica, Universidade de S\~ao Paulo, S\~ao Paulo, Brazil}

\begin{abstract}
  Recent results from the Experiment to Detect the Global Epoch of
  Reionization Signature (EDGES) show an anomalous spectral feature at
  redshifts $z\sim 15-20$ in its 21-cm absorption signal. This
  deviation from cosmological predictions can be understood as a
  consequence of physics that either lower the hydrogen spin
  temperature or increases the radiation temperature through the
  injection of soft photons in the bath. In the latter case, standard
  model neutrino decays $\nu_i \to \nu_j\,\gamma$ induced by effective
  magnetic and electric transition moments ($\mu_\text{eff}$) are
  precluded by the tight astrophysical constraints on
  $\mu_\text{eff}$. We show that if mirror neutrinos are present in
  the bath at early times, an analogous mechanism in the mirror sector
  can lead to a population of mirror photons that are then
  ``processed'' into visible photons through resonant conversion, thus
  accounting for the EDGES signal. We point out that the mechanism can
  work for mirror neutrinos which are either heavier than or
  degenerate with the standard model (SM) neutrinos, a scenario
  naturally realized in mirror twin Higgs models.
\end{abstract}

\maketitle
\section{Introduction}
\label{sec:intro}
After recombination the universe was filled with radiation, dark
matter (DM) particles and primordial gas (mainly hydrogen). This
cosmic stage, known as the ``dark ages'', lasted until the formation
of the first structures, an event that started when Compton scattering
processes could not maintain the gas and the radiation in
equilibrium. The gas, being cooled faster than the radiation field,
got gravitationally trapped in DM haloes and eventually ended up
collapsing and fragmenting, giving rise to the appearance of stars,
quasars and galaxies. At lower redshifts the Lyman-alpha photons
emitted by these first structures led to a re-ionization period, known
as the re-ionization era.

The only known observable with which the dark ages can be
observationally accessed is the 21-cm line of the ground-state
hyperfine transition of atomic hydrogen \cite{Pritchard:2011xb}. This
probe provides as well a way to test the re-ionization epoch, thus
allowing the study of the cosmic time when astrophysical objects
became the dominant source of the intergalactic medium.  The cosmic
microwave background (CMB) photons are resonantly absorbed by the
hydrogen atoms, thus producing a change in the CMB brightness
temperature $T_{21}$, which at present time depends upon cosmological
parameters, redshift and the radiation and spin temperatures
$T_\text{CMB}$ and $T_s$ (see discussion in sec. \ref{sec:general})
the latter characterizing the relative population of the hyperfine
energy levels of neutral hydrogen.  $T_s$ is determined by the
coupling with $T_\text{CMB}$ through the absorption of CMB photons as
well as by its coupling with $T_\text{gas}$ that happens through
either collisions among the hydrogen atoms or the absorption of
Lyman-alpha photons. In the absence of non-standard physics, both
$T_\text{CMB}$ and $T_s$ are well determined and so is the brightness
temperature. Observation of any deviation on this prediction can
therefore be interpreted in terms of new physics effects, for instance
non-Gaussianities \cite{Cooray:2006km,Munoz:2015eqa} or baryon-DM
interactions \cite{Tashiro:2014tsa,Munoz:2015bca}.

Recently the Experiment to Detect the Global Epoch of Reionization
Signature (EDGES) has reported on the measurement of the CMB
brightness temperature. The signal is the result of the recoupling of
$T_\text{gas}$ and $T_s$ due to Lyman-alpha photons from early stars.
The observed absorption profile is centered at around $z\simeq 17$ and
covers redshifts in the range 15-20 \cite{Bowman:2018yin}. To a large
extent, the profile is consistent with cosmological predictions, but
the observed amplitude indicates more absorption than expected. The,
arguably, most simple explanation would be an earlier
$T_\text{CMB}-T_\text{gas}$ decoupling (at $z\simeq 250$ rather than
$z\simeq 150$), which would produce an earlier cooling of the
gas. This, however, does not work since it requires the ionization
fraction to be less than the expected fraction by about an order of
magnitude, something strongly disfavored by Planck data
\cite{Bowman:2018yin}.

An explanation of the observed spectral profile requires either
decreasing $T_\text{gas}$ (gas cooling) or increasing $T_\text{CMB}$
(radiation heating). And indeed since the release of the EDGES result
both alternatives have been studied in the literature.
Ref. \cite{Barkana:2018lgd} considered DM-baryon scatterings
determined by a velocity-dependent cross section resulting from a
Coulomb-like interaction.  After discarding the possibility of a light
mediator due to fifth force constraints, ref. \cite{Berlin:2018sjs}
showed that subdominant millicharged DM can explain the 21-cm spectral
feature, despite in a constrained region in parameter space that must
be endowed with an additional depletion mechanism to prevent
overproduction.  It has been pointed out that this constraints can be
relaxed provided the millicharged DM is produced after recombination
\cite{Kang:2018qhi}.  Ref. \cite{Fraser:2018acy} discussed various
mechanisms, among which those based on the emission of soft photons
that can heat up the radiation temperature \cite{Feng:2018rje}. Using
dipole DM as a benchmark model \cite{Sigurdson:2004zp}, it ruled out
these kind of scenarios. Other mechanisms put forward include black
hole remnants from Pop-III stars \cite{Ewall-Wice:2018bzf},
interacting dark energy models \cite{Costa:2018aoy}, charge
sequestration models \cite{Falkowski:2018qdj} and more relevantly for
our study dark-photon to photon resonant conversion
\cite{Pospelov:2018kdh}. This latter relies on a non-thermal
population of dark photons, resulting from the decay of an unstable
relic, which are then resonantly converted into photons at redshifts
$z\simeq 17$.

Neutrinos can couple to electromagnetic radiation through electric
charge (milli-charged), electric/magnetic dipole (transition) moments
and/or anapole moments. Of these couplings those better understood and
probably with best experimental prospects are magnetic
dipole/transition moments $\mu_\text{eff}$ (see
e.g. ref. \cite{Giunti:2014ixa} for a review). They enable neutrino
decay processes $\nu_i\to\nu_j+\gamma$ and so ---in principle--- could
contribute to the radiation temperature at early times. However, they
contribute as well to astrophysical processes of which stellar cooling
places pretty stringent constraints on their values
$\mu_\text{eff}\lesssim 3.0\times 10^{-12}\mu_B$ (with $\mu_B$ the
Bohr magneton) \cite{Raffelt:1990pj}. In this paper, we start by
checking whether despite these bounds one could moderately raise the
radiation temperature by injecting photons though neutrino
decays. After showing that the bounds on $\mu_\text{eff}$ always lead
to a suppressed photon flux, we then entertain the possibility that
mirror neutrinos endowed with the same type of couplings can inject a
sufficiently high photon flux so to enable addressing the EDGES
anomalous spectral feature. We study in detail mirror neutrino decays
to mirror (dark) photons, $\nu'_i\to \nu_j' + \gamma\,'$, occurring at
high redshift and then getting resonantly converted into visible
photons $\gamma\,' \to \gamma$. For that aim we consider two scenarios
inspired in mirror twin Higgs models defined by degenerate and
non-degenerate SM and mirror neutrino masses with $T'<T$ (where $T'$
and $T$ refer to the mirror and SM temperatures respectively), as
required by cosmological constraints on additional dark radiation
$\Delta N_\text{eff}$ \cite{Ade:2015xua}.

The rest of the paper is organized as follows. In
sec. \ref{sec:general} we discuss generalities on the 21-cm absorption
signal and settle the conditions required for addressing the EDGES
signal. In sec.~\ref{sec:extra-rad-neutrino-decays} we consider the
case of SM neutrino decays during the redshifts relevant for EDGES, we
discuss in more detail current bounds on neutrino transition moments
and calculate the photon flux assuming $\mu_\text{eff}$ is a free
parameter. In sec. \ref{sec:mirror-world} we consider mirror neutrino
decays and resonant conversion of dark photons into visible ones. We
then provide a theoretical motivation in
sec. \ref{sec:mirror-higgs-model}, based on mirror twin Higgs models,
for the mirror neutrino scenarios we consider. In
sec. \ref{sec:conclusions} we summarize and present our conclusions.

\section{Generalities}
\label{sec:general}
During the recombination era ($z\sim 1100$) electrons and protons
recombined to form neutral hydrogen. As shown by the high degree of
isotropy of the CMB, the universe was highly uniform at that time thus
suggesting that few, if any, luminous objects could have
formed. Adiabatic expansion thus led to a stage in which the universe
consisted mainly of a neutral gas, CMB photons and DM particles, a
cosmic stage known as the dark age.  The universe evolved
adiabatically and the radiation temperature, $T_\text{CMB}$, decreased
with redshift according to $T_\text{CMB}=2.7(1+z)\,$K. The remaining
small ionization fraction, $X_e=n_e/n$, enabled the injection of
energy from the CMB to the gas through Compton scattering processes,
thus keeping both baryons and radiation at the same temperature until
$z\sim 150$.

The virial temperature of a DM halo ($T_\text{vir}$) determines the
binding energy of the material within the halo. Accordingly, only gas
for which $T_\text{gas}<T_\text{vir}$ can be trapped by the halo
gravitational pull. For $z\lesssim 150$, Compton scattering effects
became less effective and so the temperature of the gas decreased
faster than the radiation temperature. The gas then was trapped by the
DM halo, but the shocks induced by the gravitational collapse heated
up the gas to $T_\text{vir}$, thus driving the system to hydrostatic
equilibrium.  After departuring from this state, the gas contracted
within the halo and became gravitationally stable, at some point it
fragmented and led to the formation of the first stars, quasars and
galaxies. The high-energy radiation emitted from these first objects
reionized the hydrogen in the intergalactic medium, leading to the
re-ionization epoch.

The only known observable with which the dark age period can be
studied is the redshifted hydrogen hyperfine transition spectral line.
It enables as well detailed studies of the epoch of re-ionization such
as structure formation and the formation of the first galaxies.  The
ground state of neutral hydrogen is split into two hyperfine states
due to proton-electron spin-spin coupling: a singlet, corresponding to
the anti-alignment of the two spins and a degenerate triplet state
corresponding to the alignment of both spins. The energy splitting
between these states is $\Delta E=E_1-E_0\simeq 5.9\;\mu\text{eV}$,
which corresponds to a $\sim 21\,$cm photon wavelength and a
rest-frame frequency $\nu_{10}=1420\,$MHz, redshifted as
$\nu(z)=1420/(1+z)\,$MHz. Some of the CMB photons propagating in the
medium can be absorbed by hydrogen resulting in a singlet-triplet
transition which modifies the brightness temperature of the CMB
according to \cite{Madau:1996cs}
\begin{equation}
  \label{eq:brightness-CMB}
  T_b(z)=T_\text{CMB}(z)e^{-\tau(z)} + \left(1 - e^{-\tau(z)}\right)T_s(z)\ .
\end{equation}
Here $T_\text{CMB}(z)$ is the brightness temperature of the CMB
without absorption, $T_s(z)$ is the spin temperature which
characterizes the relative population of the triplet to the singlet
states.  The optical depth reads
\begin{equation}
  \label{eq:optical-depth}
  \tau(z)=\frac{3 c^2 h_{\rm P} \,A_{10}\,n_\text{HI}(z)}
  {32\pi\,\nu_{10}^2 \, k_B T_s(z)H(z)}\ ,
\end{equation}
with $A_{10}\simeq 2.9\times 10^{-15}\,\text{s}^{-1}$ the spontaneous
decay rate for the excited to the ground hyperfine states,
$n_\text{HI}$ the density of neutral hydrogen, $c$ the speed of light, 
$h_{\rm P}$ the Planck constant, $k_B$ the Boltzmann constant 
and $H(z)$ the Hubble expansion rate. 
Since $\tau\ll 1$, the change in the
brightness temperature seen today
$T_{21}(z)=(T_b(z)-T_\text{CMB}(z))/(1+z)$ can be
recast as follows
\begin{align}
  \label{brightness-temp-change}
  T_{21}\simeq \mathcal{F}
  \left(\frac{0.15}{\Omega_{m}h^{2}}\right)^{1/2}
  \left(\frac{1+z}{10}\right)^{1/2}
  \left(\frac{\Omega_{b}h^{2}}{0.02}\right)
  \left[1-\frac{T_\text{CMB}\left(z\right)}{T_s\left(z\right)}\right]\ .
\end{align}
where $\mathcal{F}=2.3\;\text{mK}\;x_\text{HI}(z)$ ($x_\text{HI}$
is the neutral hydrogen fraction), $\Omega_m$ and $\Omega_b$ are
respectively the matter and baryon energy densities in units of the
critical density and $h$ is the Hubble constant in units of 100
km/s/Mpc.

At the center of the absorption profile the redshift amounts to
$z \sim 17$. The quantities entering (\ref{brightness-temp-change})
at such redshift take the following values:
$x_{{\rm HI}}\left(17\right)\simeq 1$,
$T_s\left(17\right)=T_\text{gas}\simeq 7\,\text{K}$
\cite{Tseliakhovich:2010bj} and
$T_\text{CMB}\left(17\right)=2.7\times 18\,\text{K}\simeq
49\,\text{K}$.
Thus, the expected value for the brightness temperature contrast is
$T_{21}(z=17)=-0.2\,$K. The value provided by EDGES for the same
redshift is in contrast
$T_{21}(z=17)_\text{EDGES}=-0.5_{-0.5}^{+0.2}$K at 99\% CL, which
corresponds to about a 3.8$\sigma$ deviation from theoretical
expectations. In general $T_\text{gas}\leq T_s\leq T_\text{CMB}$, and
so the lowest spin temperature corresponds to the case
$T_s=T_\text{gas}$ (full Lyman-alpha coupling) \cite{Madau:1996cs},
thus representing the case where $T_\text{CMB}/T_s$ is the
largest. Under this assumption, the EDGES signal can be reconciled if
the ratio $T_\text{CMB}/T_s$ is enhanced by a factor $2$. If one
departures from this assumption and considers the more general case
where $T_s>T_\text{gas}$ this factor should increase accordingly
\cite{Venumadhav:2018uwn}. For concreteness throughout our analysis we
will assume $T_s=T_\text{gas}$.

\section{Extra radiation from
  electromagnetic-induced neutrino decays}
\label{sec:extra-rad-neutrino-decays}%
The injection of soft photons in the early universe by the SM neutrino
decays can proceed through magnetic (electric) transition moments
$\mu_{ij}$ ($\epsilon_{ij}$). In the following we consider
the effective electromagnetic neutrino interactions\footnote{In the
  rest of the article, we will use natural units
  $c = h_{\rm P}/(2\pi) = k_B = 1$.}
\begin{equation}
\label{eq:eff-Lag}
{\cal L}_\text{eff}= \frac{1}{2}\bar{\nu}_{i}\sigma_{\mu\nu}
  \left(\mu_{ij}+\epsilon_{ij}\gamma_{5}\right)\nu_{j}F^{\mu\nu},
\end{equation}
where $\sigma_{\mu\nu}=i[\gamma_{\mu},\gamma_{\nu}]/2$, $F^{\mu\nu}$
is the electromagnetic field strength tensor and $i, j$ label neutrino
mass eigenstates. Note that here we have assumed neutrinos are Dirac
particles, assuming otherwise will not change our conclusion. These
couplings induce radiative neutrino decays $\nu_i\to\nu_j+\gamma$ for
which the decay width can be written as
\begin{equation}
  \label{eq:decay-width}
  \Gamma_{\nu_{i}\to\nu_{j}+\gamma} = \frac{\mu_{\text{eff},ij}^{2}}
  {8\pi}\left(\frac{\Delta m_{ij}^{2}}{m_i}\right)^{3} \ ,
\end{equation}
with $\mu_{\text{eff},ij}=\sqrt{|\mu_{ij}|^2 + |\epsilon_{ij}|^2}$,
$\Delta m_{ij}^{2}\equiv m_{i}^{2}-m_{j}^{2}$ and $m_i$ the $i$-th
neutrino mass eigenstate. 
(Where family index is not relevant, neutrino mass will be denoted 
simply as $m_\nu$.)
This effective coupling is subject to tight
constraints from laboratory experiments and astrophysical
considerations (see e.g. \cite{Giunti:2014ixa}). For the former, the
most severe bound is derived from the GEMMA experiment which relies on
measurements of electron recoils induced by the neutrino-electron
elastic scattering process $\nu e \to \nu e$ \cite{Beda:2010hk}. The
current 90\% CL limit neglecting atomic effects reads (from now on we
will drop family indices, except in those cases where strictly
necessary)
\begin{equation}
  \label{eq:Gemma}
  \mu_\text{eff}\lesssim 3.2\times 10^{-11}\,\mu_B\ .
\end{equation}
Astrophysical bounds are
more stringent, in particular those derived from plasmon decay
($\gamma\to \nu\nu$) in globular cluster stars
\cite{Raffelt:1990pj}. This process---enabled by medium
effects---releases an amount of energy through the neutrinos that
escape the stellar medium, resulting in a delay in helium ignition and
thus in the following upper limit
\begin{equation}
  \label{eq:upper-bound-astrophysical}
  \mu_\text{eff}\lesssim 3.0\times 10^{-12}\,\mu_B\ .
\end{equation}

Next, notice that photons injected much before recombination
($z\sim 1100$) will get fully absorbed by the plasma, while those
injected below $z\sim 15$ cannot contribute to the spectral distortion
observed by EDGES. Thus, if photons emitted in neutrino decays were to
be responsible for the EDGES signal, they should be generated in the
window $15 \lesssim z\lesssim 1100$ with energy falling within the
EDGES energy absorption interval $[0.28,0.37]\,\mu\text{eV}$
(rest-frame frequency redshifted in the interval $15<z<20$). To
determine which photons can contribute to the signal, one needs their
energy at production properly redshifted, \textit{c'est-\`a-dire}
$E(z)=\Delta m_{ij}^2/2/m_i/(1+z)$, assuming decays at rest.  The
condition of this energy falling within the absorption energy range,
fixes the minimum (and maximum) mass that the decaying neutrinos
should have so to be ``visible''. Using the best fit point values for
the mass squared differences \cite{deSalas:2017kay}, we find
$m_h\gtrsim 3\,$eV ($h=3$ for normal order neutrino mass spectrum,
$h=1$ for inverted order) and $m_2\gtrsim 0.1\,$eV.  Cosmological
constraints on neutrino masses $\sum_i m_i \le 0.68\,$eV (95\% CL
limit) \cite{Ade:2015xua}, thus imply that photons produced by the
decay of $\nu_h$ will fall outside the EDGES energy window and only
decay of $\nu_2$ matters.
\begin{figure*}[t]
  \centering
  \includegraphics[scale=0.35]{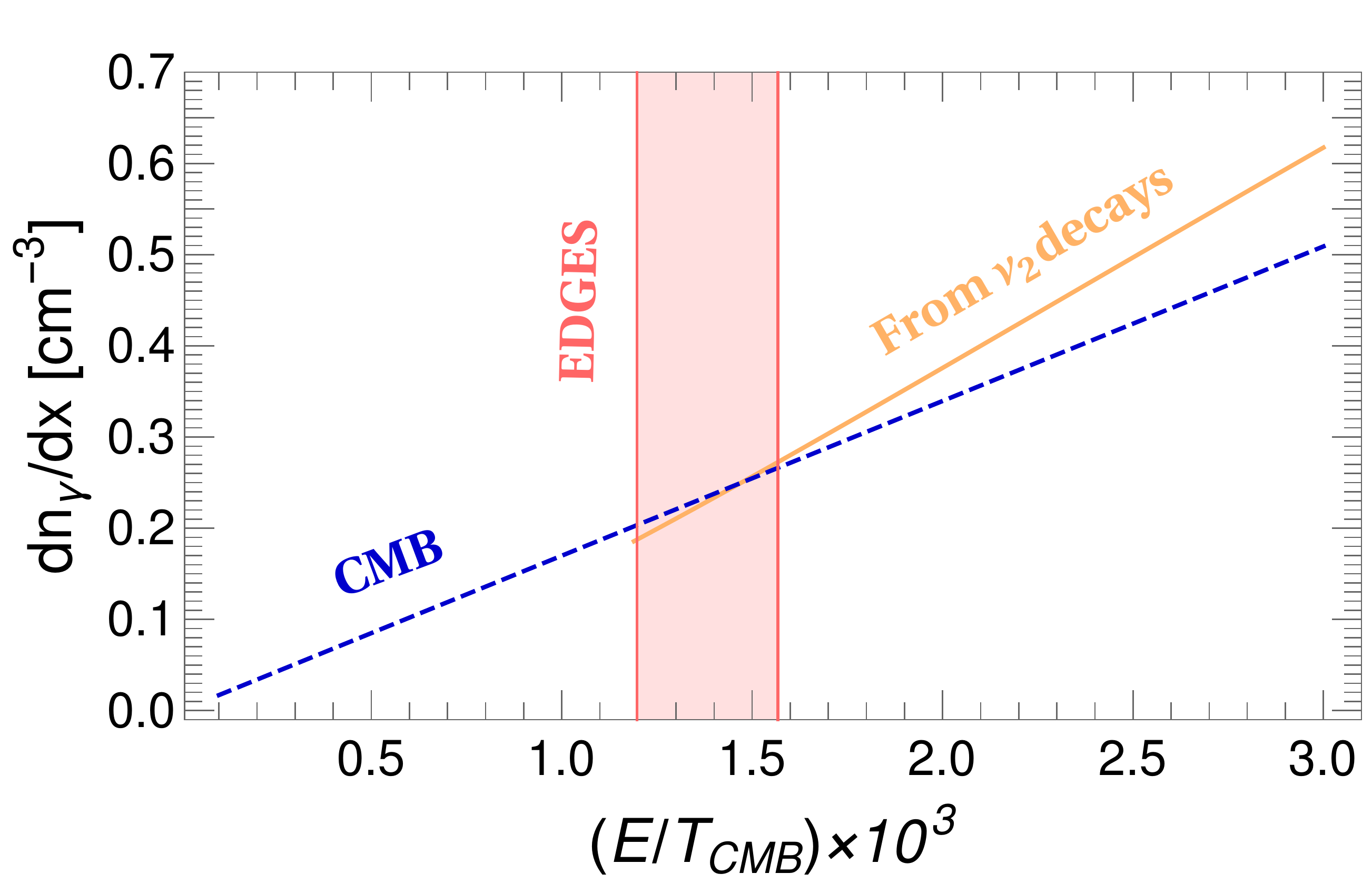}
  \caption{Resulting photon spectrum from radiative neutrino decays
    (orange solid curve) compared to that of the CMB (blue dashed
    line), $x=E/T_\text{CMB}$. The shaded area indicates the EDGES
    absorption frequency window $[68,89]\,$MHz. This result has been
    derived by requiring the extra radiation to amount to that of the
    CMB, as required by the EDGES signal under the assumption of full
    Lyman-alpha coupling \cite{Bowman:2018yin}. The cutoff to the left
    in the photon spectrum from $\nu_2$ decays, corresponds to
    redshifts for which $z\gtrsim 1100$.}
  \label{fig:SM-neutrino}
\end{figure*}
To determine the contribution of neutrino decays to the number of
photons in the plasma, one should calculate the photon number density
per-unit energy, which at present time reads \cite{Masso:1999wj}
\begin{equation}
  \label{eq:photon-number-density}
  \frac{dn_{\gamma}}{dE} = \frac{B}{E}\frac{n_{\nu}\left(t_0\right)
    \langle \Gamma_{\nu}\rangle}{H\left(z\right)}
  e^{-\langle\Gamma_{\nu}\rangle\,t\left(z\right)} \ ,
\end{equation}
where $E$ is the photon energy today, $B$ the branching fraction for
the radiative decay, and $n_{\nu}\left(t_{0}\right)$ the would-be
present number density of neutrinos if they did not decay. Here
$\langle\Gamma_{\nu}\rangle$ is the thermally averaged total decay
width of neutrino $\nu$
\begin{equation}
  \label{eq:thermally-averaged-xsec}
  \langle\Gamma_{\nu}\rangle=
  \Gamma_{\nu}^0\,\frac{K_1(m_\nu/T_\text{CMB})}{K_2(m_\nu/T_\text{CMB})}\ ,
\end{equation}
with $\Gamma_{\nu}^0$ the temperature-independent total decay width
and $K_{i}(x)$ the order $i$-th modified Bessel function of the second
type.  The expansion time $t\left(z\right)$ is given in terms of the
Hubble expansion rate $H\left(z\right)$, namely
\begin{equation}
  \label{eq:exp-time}
  t\left(z\right) = \int_{z}^{\infty}
  \frac{dz'}{1+z'}\frac{1}{H\left(z'\right)} \ ,
\end{equation}
which for a flat universe with matter $\Omega_{m}$ and a non-vanishing
cosmological constant $\Omega_{\Lambda}$, is given by\footnote{We have
  included the radiation contribution $\Omega_{r}=5.38\times10^{-5}$
  \cite{Patrignani:2016xqp} which is negligible for $z\lesssim1100$
  but will be important for our considerations in the next section.}
$H\left(z\right)=H_{0}\sqrt{\Omega_{\Lambda}
  +\Omega_{m}\left(1+z\right)^{3}+\Omega_{r}\left(1+z\right)^{4}}$.
For our calculation we have taken $H_{0}=67.8$ km/s/Mpc,
$\Omega_{\Lambda}=0.69$ and $\Omega_{m}=0.31$
\cite{Patrignani:2016xqp}. Bearing in mind that the would-be number
density of SM neutrinos per generation today is
\begin{equation}
  \label{eq:neutrino-density}
  n_{\nu}\left(t_{0}\right) = \frac{3}{2}
  \frac{\zeta\left(3\right)}{\pi^{2}}\left(\frac{4}{11}\right)
  T_{0}^{3} \ ,
\end{equation}
where $T_{0}=2.725$ K is temperature of the CMB photons today, the
contribution to the photon number density from $\nu_2$ decays can then
be calculated. Assuming full Lyman-alpha coupling ($T_\text{gas}=T_s$)
\cite{Bowman:2018yin}, this contribution should amount to that of the
CMB so to account for the EDGES signal (see sec.~\ref{sec:general}).
Fig. \ref{fig:SM-neutrino} shows the result for the photon number
density due to $\nu_2$ radiative decays in comparison to the
Rayleigh-Jeans tail of the CMB black body spectrum.  This result has
been derived assuming a normal order neutrino mass spectrum (inverse
mass ordering gives similar results), using $m_2=0.12\,$eV which
minimizes the required $\mu_\text{eff}=7.8\times 10^{-6}\mu_B$, the
latter a value far larger than current limits.  Thus, an explanation
of the anomalous spectral distortion observed by EDGES, based on
electromagnetic-induced radiative neutrino decays requires effective
electromagnetic couplings already ruled out by data.

It is worth pointing out that even if one could afford a sufficiently
large $\mu_\text{eff}$, there are extra effects one should deal
with. First of all since $\Gamma_{\nu_3}>\Gamma_{\nu_2}$, $\nu_3$
decays will yield a more abundant photon flux in the energy range
$\sim[2,125]\,\mu\text{eV}$ and will contribute sizeably to the CMB at
redshifts above the EDGES window. That effect, however, could be kept
under control by assuming that the effective transition magnetic
moments of $\nu_3$ are suppressed. As can be seen in
fig. \ref{fig:SM-neutrino} one finds the same effect for $\nu_2$. And
of course in this case the solution used for $\nu_3$ will not work,
implying that the scenario will be further constrained from
measurements of distortions to the CMB at redshifts $z\gtrsim 30$.
\section{Extra radiation from a mirror sector}
\label{sec:mirror-world}
The conclusion reached in the previous section might change if mirror
neutrinos couple to radiation in the same way SM neutrinos do. Let us
discuss this scenario in more detail. The electromagnetic couplings of
the mirror neutrinos resemble those in (\ref{eq:eff-Lag}), with
neutrinos and the electromagnetic field tensor traded for those of the
mirror sector, which we will denote as $\nu^\prime$ and
$F_{\mu\nu}^\prime$. For the coupling we will use
$\mu_\text{eff}^\prime=\sqrt{|\mu'_{ij}|^2 + |\epsilon'_{ij}|^2}$,
with $\mu'_{ij}$ and $\epsilon'_{ij}$ the mirror neutrino magnetic and
electric transition moments. In addition to these couplings one has as
well a kinetic mixing term which couples the electromagnetic field
tensors of the visible and mirror sectors,
$(\varepsilon/2) F^{\mu\nu}F_{\mu\nu}^\prime$. The simultaneous
presence of $\mu_\text{eff}^\prime$ and $\varepsilon$ induces
processes of the type $\gamma\to \nu^\prime\nu^\prime$, which as in
the SM case leads to stellar cooling and thus to the upper limit
\begin{equation}
  \label{eq:stellar-cooling-mirror}
  \varepsilon\,\mu_\text{eff}^\prime\lesssim 3.0\times 10^{-12}\mu_B\ .
\end{equation}

The mirror sector is subject as well to cosmological constraints which
require the SM temperature to be larger than the mirror sector
temperature. This can be understood from the contribution of mirror
neutrinos to the effective ``neutrino'' degrees of freedom
\begin{equation}
  \label{eq:mirror-sector-Neff}
  \Delta N_\text{eff}=\frac{4}{7}
  \left(\frac{11}{4}\right)^{4/3}\,g_\star^\prime\ ,
\end{equation}
where $g_\star^\prime$ refers to the effective relativistic degrees of freedom
which is given by
\begin{equation}
  \label{eq:relativistic-dof}
  g_{\star}^\prime=\sum_{i=\text{boson}}
  g_{i}'\left(\frac{T_{i}'}{T_{\gamma}}\right)^{4}
  +
  \frac{7}{8}\sum_{i={\text{fermion}}}g_{i}'
  \left(\frac{T_{i}'}{T_{\gamma}}\right)^{4}\ .
\end{equation}
Here $T_\gamma$ is the photon temperature while $T'_i$ is the
temperature of the corresponding mirror sector relativistic degree of
freedom. Assuming $T_i'=T'$ (common temperature for all mirror sector
relativistic degrees of freedom) and that at the time of $\nu_i'$
decay only the dark photon and the three mirror neutrino species are
relativistic, $\Delta N_\text{eff}$ becomes
\begin{equation}
  \label{eq:Delta_Neff}
  \Delta N_{\text{eff}} = \frac{29}{7}
  \left(\frac{11}{4}\right)^{4/3}
  \left(\frac{T'}{T_{\gamma}}\right)^{4}\ .
\end{equation}
Thus, by using the $2\sigma$ limit $\Delta N_{\text{eff}}<0.65$ from Planck
(the bound from Big Bang nucleosynthesis (BBN) is comparable) \cite{Ade:2015xua} 
an upper bound on the temperature of the mirror
sector can be derived, $T'<0.45 T_\gamma$. 
This bound could be relaxed
if the heaviest and next-to-heaviest mirror neutrinos decay before BBN
($z\sim 4\times 10^8$) and the lightest one is stable, in that case
the mirror sector can be slightly hotter, $T'<0.53\,T_\gamma$.

Another bound one has to consider has to do with mirror neutrino
masses $m_{\nu'} = \{m_i'\}$, which can be constrained by combining
(\ref{eq:Delta_Neff}) with cosmological limits on $\sum_i m_i$ (see
sec. \ref{sec:extra-rad-neutrino-decays}). Taking
$m_{\nu'}=r\, m_\nu$, with $r$ a common rescaling that determines how
heavy the mirror neutrinos can be, and using
$\sum_i m_i + \sum_i m_i' (n_\nu'/n_\nu) < 0.68\,$ eV
\cite{Ade:2015xua} we find the following upper limit
\begin{align}
  \label{eq:contribution-to-meff}
  r\lesssim 55 \,\left(\frac{0.65}{\Delta N_\text{eff}}\right)^{3/4}
  \left[
  \left(\frac{0.05\,\text{eV}}{\sum_i m_i}\right) - \frac{5}{68}
  \right]\ .
\end{align}
Here we normalize the sum of SM neutrino masses to the
value of the atmospheric mass scale determined from neutrino
oscillation data \cite{deSalas:2017kay}.
\begin{figure*}[t]
  \centering
  \includegraphics[scale=0.35]{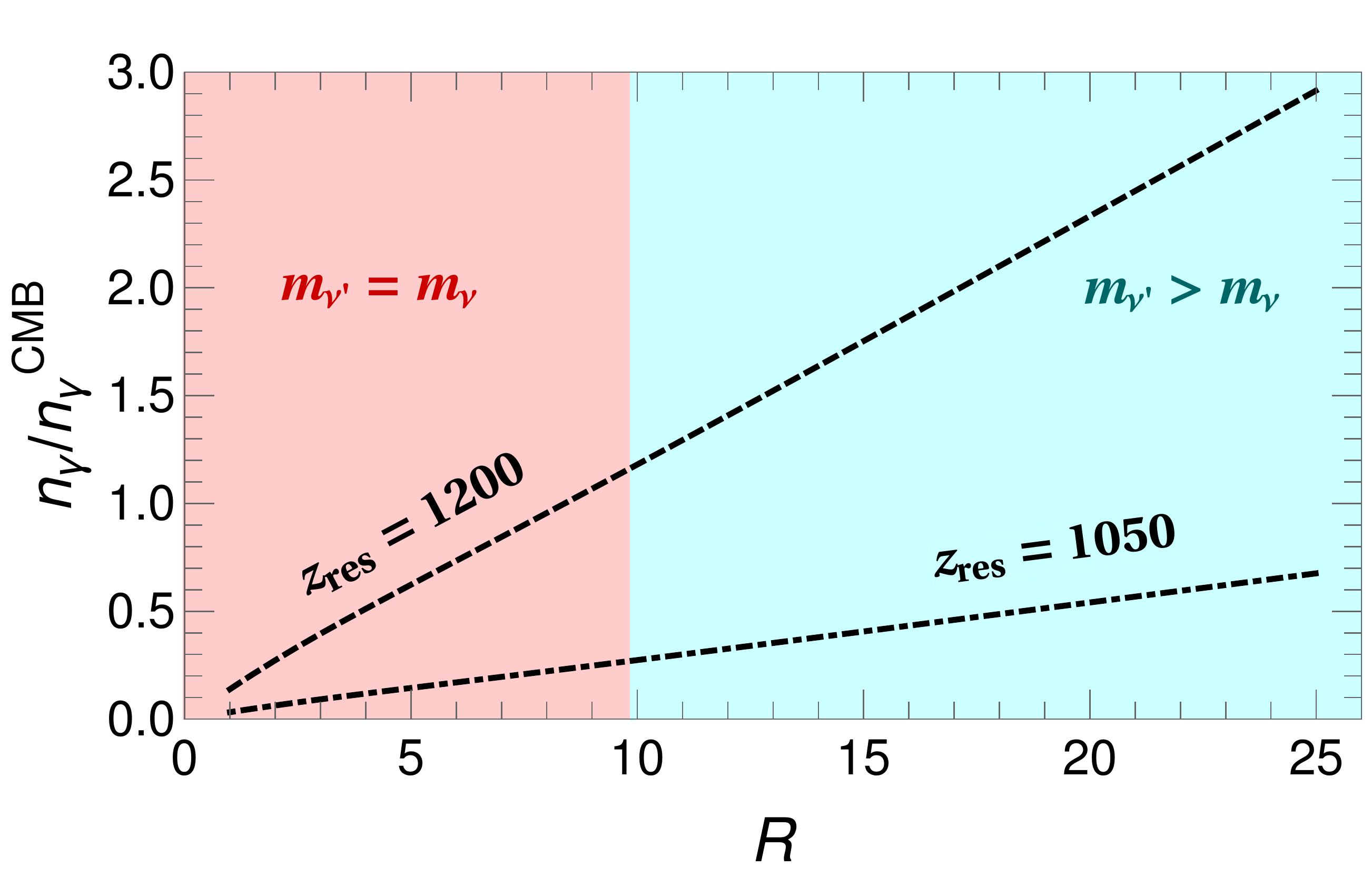}
  \caption{Extra radiation from mirror neutrino decay to dark photons
    and then processed to visible radiation through $\gamma\,'-\gamma$
    conversion as a function of $R=(m_3'/0.05 \text{eV})$ (assuming a
    normal order mass spectrum). The temperature of the mirror sector
    has been fixed to $T'=0.4\,T_\gamma$ and
    $n_\gamma=n_\gamma^{\prime\text{CMB}}+n_\gamma^\text{decay}$
    ($n_\gamma^{\prime\text{CMB}}$ refers to the amount of dark
    CMB$^\prime$ radiation converted into visible radiation).  We have
    chosen two benchmark values of the redshift for which resonant
    conversion occurs at $z_\text{res}=1200$ and $1050$ which
    correspond to
    $m_\gamma\simeq m_{\gamma\,'}\simeq 4.4\times 10^{-10}\,$eV and
    $2.1\times 10^{-10}\,$eV, respectively.  The vertical line which
    separates the two shaded regions corresponds to
    $\sum_i m_i + \sum_i m_i' (n_\nu'/n_\nu) = 0.68\,$ eV and in the
    left (right) region, the SM and mirror neutrinos can (cannot) be
    degenerate.  }
  \label{fig:mirror-case}
\end{figure*}
\subsection{Dark photon resonant conversion}
\label{sec:dark-photon-res-conv}
Mirror neutrino decays can directly generate a photon flux through
kinetic mixing, $\nu_i'\to\nu_j'+\gamma$. These decays however will be
controlled by $\varepsilon\,\mu'_\text{eff}$, and so given the bound
in (\ref{eq:stellar-cooling-mirror}) the photon flux will be rather
suppressed (pretty much resembling what we found in the SM neutrino
case). On the other hand, mirror neutrinos decay to dark photons
($\nu_i'\to\nu_j'+\gamma\,'$) can yield a population which is not
necessarily small. The key point is that these decays are solely
determined by $\mu_\text{eff}'$, which can be large if
$\varepsilon \ll 1$ while satisfying
(\ref{eq:stellar-cooling-mirror}).  These decays can take place way
above $z\sim 1100$, as far as they occur after the mirror and SM
sectors have thermally decoupled, to avoid $\gamma\,'-\gamma$
thermalization.

Once in the bath, as we will see shortly, depending on the mass of
dark photons, they can be efficiently ``processed'' into visible
photons through resonance conversion, even with small $\varepsilon$
enforced by (\ref{eq:stellar-cooling-mirror}) together with the CMB
constraints \cite{Mirizzi:2009iz}.  In contrast to mirror neutrino
decays, the conversion process should occur in the window
$15\lesssim z\lesssim 1100$ if this mechanism is to explain the
anomalous spectral profile reported by EDGES for the reasons
elaborated after eq. \ref{eq:upper-bound-astrophysical}.

In the heat bath visible photons acquire an effective mass through the
scattering with free electrons and neutral atoms. Neglecting the
latter it can be written according to \cite{Mirizzi:2009iz}
\begin{equation}
  \label{eq:eff-mass}
  m_\gamma\simeq 1.75\times 10^{-14}\,(1+z)^{3/2}\,X_e^{1/2}\,\text{eV}\ ,
\end{equation}
where the free electron fraction $X_e$ can be well approximated for
$z\gtrsim 70$ by the expression \cite{Mirizzi:2009iz}
\begin{equation}
  \label{eq:ionization-fraction-apprx}
  \log X_e\simeq -\frac{3.15}{1 + e^{\bar z}}\qquad
  (\bar z=\frac{z-907}{160})\ .
\end{equation}
Resonant $\gamma\,'-\gamma$ conversion, which resembles the MSW effect
for solar neutrinos
\cite{Wolfenstein:1977ue,Mikheev:1986gs,Mikheev:1986wj}, happens when
the dark photon mass amounts that of the visible photon mass,
$m_\gamma\simeq m_{\gamma\,'}$. In that case the $\gamma\,'-\gamma$
conversion probability can be taken as
\cite{Parke:1986jy,Mirizzi:2009iz}
\begin{equation}
  \label{eq:conversion-prob}
   P_{\gamma\,'\to\gamma} = P_{\gamma\to\gamma\,'} 
   \simeq 1 - e^{-2\pi\, r\, k \sin^2\varepsilon}\ ,
\end{equation}
which holds for $\varepsilon \ll 1$.  The second term corresponds to
the level crossing probability with $k=m_{\gamma\,'}^2/(2 E)/(1+z)$
and
$r=|d\log m^2_\gamma/dt|^{-1}_{t=t_\text{res}} = |d\log
m^2_\gamma/dz|^{-1}/(1+z)/H(z)|_{z=z_\text{res}}$.
The setup of eqs. (\ref{eq:eff-mass}) and
(\ref{eq:ionization-fraction-apprx}) as well as the definitions for
the parameters $k$ and $r$ allow the determination of
$P_{\gamma\,'\to\gamma}$ and therefore of the corresponding photon
spectrum
\begin{equation}
  \label{eq:photon-spectrum-mirror}
  \frac{dn_{\gamma}}{dE} = P_{\gamma\,'\to\gamma}\times
  \frac{B}{E}\frac{n_{\nu'}\left(t_{0}\right)\langle\Gamma_{\nu'}\rangle}
  {H\left(z\right)}e^{-\langle\Gamma_{\nu'}\rangle t \left(z\right)}
  \,\theta\left(z-z_\text{res}\right)\ ,
\end{equation}
where the Heaviside function assures that dark photons produced below
$z_\text{res}$ (the redshift for which the resonance occurs) will not
be converted into visible photons. To show that this mechanism can
account for the EDGES signal, we fix
$\mu_\text{eff}'=3\times 10^{-5}\,\mu_B$, $\varepsilon=10^{-7}$ and
$T'/T_\gamma=0.4$.  For illustration, we have chosen the redshifts for
which resonance conversion occurs to be $z_\text{res}=1200$ and
$1050$.  These fix $m_{\gamma\,'} \simeq m_\gamma$ to be
$4.4\times 10^{-10}\,$eV and $2.1\times 10^{-10}\,$eV respectively.
The $\varepsilon$ has been chosen such that it is consistent with the
bounds from distortions of the CMB spectrum
$\varepsilon \lesssim 10^{-6}$ \cite{Mirizzi:2009iz}.  We then
calculate the photon number density generated through
$\gamma\,'-\gamma$ conversion as a function of the heaviest mirror
neutrino mass $m_{3'}$ assuming normal order. We include photons from
mirror neutrino decays as well as from the dark background radiation
($n_\gamma=n_\gamma^\text{decay} + n_\gamma^{\prime\text{CMB}}$), and
compare with the CMB photon number density
($n_\gamma^\text{CMB}$). The result is shown in
fig.~\ref{fig:mirror-case}, where we have specified two scenarios for
the mirror neutrino mass spectrum\footnote{Although we have chosen the
  normal order neutrino mass spectrum, inverse order does not change
  our results.}: (a) complete degeneracy between SM and mirror
neutrinos, (b) non-degeneracy, $m_{\nu'}=r\,m_\nu$, with $r$ subject
to the bound in (\ref{eq:contribution-to-meff}). This result shows
that one can address the EDGES spectral feature by means of this
mechanism.
%
\section{Realization in the mirror twin Higgs model}
\label{sec:mirror-higgs-model}
The mirror neutrinos we have considered in the previous section are
naturally realized in twin Higgs models \cite{Chacko:2005pe}. In a
rather simple realization, one can understand them as a class of
models in which the scalar potential features a global $U(4)$
symmetry. The scalar field of the theory, $\cal H$, transforms as the
fundamental representation of the global symmetry group and acquires a
vacuum expectation value (vev) $v'$ that spontaneously breaks $U(4)$
to $U(3)$ leaving behind---in the absence of quantum
corrections---seven Nambu-Goldstone bosons (NGBs).\footnote{In fact,
  the scalar potential has an enhanced global symmetry $O(8)$ and one
  can consider the breaking as $O(8) \to O(7)$ which can contain the
  custodial symmetry of the SM.}  $\cal H$ is constructed out of two
doublets belonging to the gauged direct product sub-group
$SU(2) \times SU(2)'\subset U(4)$, $H$ and $H'$ (one would identify
$SU(2)$ with the SM $SU(2)_L$ and $H$ with the SM Higgs doublet).
Since the global symmetry is explicitly broken by gauge
$SU(2)\times SU(2)'$, one would expect the gauge quantum corrections
to lift the mass of the NGBs.  On the one hand, the presence of a
$\mathbb{Z}_2$ ``twin'' symmetry (that interchanges
$H\leftrightarrow H'$) leads to self-energy one-loop corrections that
are $U(4)$ invariant, and so do not contribute to
$m_\text{NGB}$\footnote{This is a consequence of $g=g'$ (with $g,g'$
  the $SU(2) \times SU(2)'$ gauge couplings respectively), implied by
  the discrete twin symmetry. Ultimately due to this symmetry the NGBs
  are insensitive to quadratic divergences.}.  On the other hand,
gauge one-loop quantum corrections to the scalars ($H$ and $H'$)
four-point functions, instead, explicitly break $U(4)$, thus implying
$m_\text{NGB}\neq 0$ (the NGBs are actually pseudo-NGBs).  These
corrections, however, are logarithmically divergent and so allow
$\mathcal{O}(m_\text{NGB})\sim g^2 v'/(4\pi)$ to be at the weak scale
for cutoff scales up to $\sim$ 5-10 TeV. Thus, identifying the SM
Higgs among these degrees of freedom prevents the Higgs mass from
acquiring large quantum corrections and therefore solves the little
hierarchy problem, with new particles which are singlets under the SM.

In the limit of exact $\mathbb{Z}_2$ symmetry both the SM Higgs vev,
$v$, and $v'$ are equal.  A mechanism that enables a mild hierarchy
between $v$ and $v'$ relies on the introduction of a term in the
scalar potential that softly breaks $\mathbb{Z}_2$ and leads to $v<v'$
\cite{Chacko:2005pe,Barbieri:2005ri}.  This small $\mathbb{Z}_{2}$
breaking is required to obtain the correct electroweak breaking scale
and Higgs precision measurements further require $v'/v \gtrsim 3$
\cite{Barbieri:2016zxn}.  Complete models can be constructed with the
aid of this mechanism, by extending the symmetry to all the
interactions of the SM or by identifying the twin symmetry with
parity, in which case two models are possible: mirror twin Higgs
models \cite{Chacko:2005pe} or left-right symmetric twin Higgs models
\cite{Chacko:2005un}. Is within the former---in which there is a
mirror copy of the SM with the same (mirror) particle content and
interactions---that the scenarios we have pointed out in the previous
section emerge and what we will focus on next. (Here $\mathbb{Z}_2$ is
the symmetry which interchanges between particles and mirror
particles.)

In the mirror twin Higgs models, further call for $\mathbb{Z}_2$
breaking arise from the constraints on extra radiation generated by
the mirror sector\footnote{Another way out is to remove the
  troublesome light mirror particles i.e. by having an imperfect copy
  of the SM in the mirror sector \cite{Craig:2015pha}.}
\cite{Barbieri:2016zxn}.  At $T\gtrsim$ GeV, both the SM and mirror
sectors are thermally coupled through Higgs-exchanged processes. As
the temperature decreases, mirror particles inject dark radiation in
the bath (in the form of light degrees of freedom) through decay and
annihilation processes.  Their contribution to dark radiation is
determined by the decoupling temperature $T_d$, below which SM and
mirror sector interactions are slower than the Hubble expansion rate
and both sectors decouple.

In the case of $\mathbb{Z}_2$-symmetric Yukawa couplings, fermion
masses in both sectors differ only by the ratio $v/v'$. And
consistency with cosmological bounds on $\Delta N_\text{eff}$ demands
this ratio to be rather large $v/v' \gtrsim 40$, implying a
fine-tuning to get the correct electroweak scale
\cite{Barbieri:2016zxn}. Breaking $\mathbb{Z}_2$ in the Yukawa sector
allows for $y_{f'}>y_{f}$ and so for heavier mirror charged leptons
and quark masses\footnote{To avoid destabilizing the Higgs mass
  $\mathbb{Z}_2$ breaking should assure $y_{t'} \simeq y_t$ holds at
  the percent level.}, which in particular imply higher mirror QCD
phase transition temperature $T'_{\text{QCD}'}$ (up to $\sim 3\,$GeV
\cite{Barbieri:2016zxn}). The amount of dark radiation can then be
reduced by having $T_d$ below $T'_{\text{QCD}'}$ (so light mirror
quarks will not contribute at decoupling) and above the SM one
$T_\text{QCD} \sim 0.2\,$GeV\footnote{Ref.  \cite{Csaki:2017spo}
  proposed to achieve $T_{\text{QCD}} < T_d < T'_{\text{QCD}'}$ by
  having significant mixing between GeV scale mirror neutrinos and the
  SM neutrinos.  Note that this scenario will not work for us since
  the number density of GeV scale mirror neutrinos will be too
  suppressed after decoupling, leading to a suppressed mirror photon
  and hence photon flux.}.  Assuming separate entropy conservation in
the mirror and SM sectors below $T_d$, their temperatures are related
through \cite{Feng:2008mu}
\begin{equation}
  \label{eq:mirror-T-twin-Higgs}
  T' = \left[
    \frac{g_{\star}\left(T\right)}{g_{\star}'\left(T^{'}\right)}
    \frac{g_{\star}'\left(T_{d}\right)}{g_{\star}\left(T_{d}\right)}
  \right]^{1/3}\,T \ .
\end{equation}
For instance, assuming that at $T_d \sim $ GeV, what remains are only
$\gamma\,'$ and $\nu'$ in the mirror sector, $g_\star'(T_d) = 7.25$,
while in the SM sector $g_\star(T_d) = 61.75$.  At
$T = T_\gamma < m_e$, $g_\star(T)=3.9$ while $g_\star'(T') = 7.25$,
thus resulting in $T' = 0.4\,T_\gamma$.  To achieve the above
temperature difference, alternatively, ref. \cite{Chacko:2016hvu}
proposed to heat up the SM sector with respect to the mirror sector by
having GeV right-handed neutrinos which decay preferentially to SM
particles.  In sec. \ref{sec:dark-photon-res-conv}, we have taken
$T' = 0.4 T_\gamma$ which is consistent with the bound on
$\Delta N_\text{eff}$ (see eq. (\ref{eq:Delta_Neff}) and the
discussion below it).

In mirror twin Higgs models, neutrino masses can be generated from
dimension five effective operators, with $\mathbb{Z}_2$ symmetric
couplings, generated from a seesaw mechanism \cite{Barbieri:2016zxn}.
The effective Lagrangian reads
\begin{equation}
  \label{eq:Lag-nu-Twin-Higgs}
  \mathcal{L}_\text{eff} = \frac{y_D}{M_1}
  \left(LH\right)^{2}+\frac{y_D}{M_1}
  \left(L'H'\right)^{2}+\frac{\lambda}{M_2}
  \left(LH\right)\left(L'H'\right) \ ,
\end{equation}
where $L$ and $L'$ are respectively the SM lepton and mirror lepton
doublets.  With $M_2\gg M_1$, neutrinos acquire mostly Dirac masses
after $H$ and $H'$ acquire vevs. Due to the $\mathbb{Z}_2$ symmetry,
$m_{\nu'}$ and $m_\nu$ differ only by $v'/v$,
\begin{equation}
  \label{eq:neutrino-masses-Dirac}
  m_{\nu'}=\left(\frac{v'}{v}\right)^2\,m_\nu\ .
\end{equation}
If on the contrary $M_1\gg M_2$, neutrinos will be Majorana with their
masses generated by the last term in~(\ref{eq:Lag-nu-Twin-Higgs}) and so
\begin{equation}
  \label{eq:nu-masses-scenario1}
  m_{\nu'} = m_{\nu}\ .
\end{equation}
Cases (\ref{eq:neutrino-masses-Dirac}) and
(\ref{eq:nu-masses-scenario1}) correspond to the two mass spectra we
have considered in our analysis.

As a final remark, in our scenario, the mirror photon should acquire a
small mass of the order of $10^{-10}$ eV. The broken mirror QED can be
achieved by having a soft $\mathbb{Z}_2$ breaking mass for the mirror
hypercharge gauge boson \cite{Chacko:2005pe}.  Furthermore, the
required small kinetic mixing $\varepsilon \lesssim 10^{-6}$ in our
scenario also implies the existence of millicharged (mirror) particles
where the current constraints on them with mass $\gtrsim 10^{-2}$ GeV
is rather loose $\varepsilon \lesssim 10^{-4}$\cite{Essig:2013lka}.
\section{Conclusions}
\label{sec:conclusions}%
In this letter we have entertained the possibility that the anomalous
spectral feature recently reported by EDGES arises from extra
radiation (soft photons) injected in the bath at early times. We first
have considered SM neutrino decays induced by neutrino magnetic and electric
transition moments $\mu_\text{eff}$, $\nu_i\to \nu_j+\gamma$.
Treating $\mu_\text{eff}$ as a free parameter and assuming full
Lyman-alpha coupling, we calculated the photon flux required to
address the EDGES signal. We find that an explanation based on
electromagnetic-induced neutrino decays requires values for
$\mu_\text{eff}$ already ruled out by stellar cooling considerations,
thus ruling out such possibility.

We have shown that in the presence of mirror neutrinos (as expected
e.g. in mirror twin Higgs models),
a larger mirror transition moment allows a sufficiently large photon
flux that can account for the EDGES signal, while simultaneously
satisfying astrophysical and laboratory bounds on $\mu_\text{eff}'$
and kinetic mixing $\varepsilon$.  The mechanism generates the
appropriate amount of radiation in a two-step process. In a first
stage mirror neutrino decays $\nu_i'\to \nu_j'+\gamma\,'$ populate the
bath with a dark photon density. The decays can occur way before
recombination provided they happen after the SM and mirror sectors
have decoupled. In a second stage, the dark photon population is
processed into visible radiation through resonant $\gamma\,'-\gamma$
conversion. The resulting additional photon density thus arises from
the mirror neutrino decays (after conversion) and the subdominant
mirror neutrino background (CMB$^\prime$).  In contrast to the first
stage, the second should take place near or after recombination, to
avoid the resulting photons from being totally absorbed by the
medium. Assuming as well full Lyman-alpha coupling we have explicitly
shown that this mechanism can raise the radiation temperature at the
levels required by the EDGES anomalous spectral feature.

Finally, we showed that the scenarios we considered can be realized
naturally in mirror twin Higgs models. Thus, on top of the
phenomenology which come along with them, we have provided another
avenue to probe them during the cosmic dark ages.
%
\section*{Acknowledgement}
DAS is supported by the grant ``Unraveling new physics in the
high-intensity and high-energy frontiers'', Fondecyt No 1171136.
CSF is supported by the Brazilian National Council for Scientific 
and Technological Development (CNPq) grant 420612/2017-3. 
He would like to thank Paola Arias Reyes for kind invitation to 
deliver lectures at the ``Cosmolog\'{i}a de Part\'{i}culas'' school at 
University of Santiago of Chile and the hospitality of 
the Federico Santa Maria Technical University at Santiago 
where this work was initiated.

\bibliography{ref}
\end{document}